\documentclass{article}
\usepackage{hiph-preprint}
\usepackage{graphicx}
\volnumber{22} \issuenumber{1} \edyear{2005}                             
\frompage{000} \topage{000}                                              
\recrevdate{18 November 2005}                                            
\def\rightmark{System size dependence of charged hadron spectra}
\def\leftmark{E. Wenger (PHOBOS)}
 
\title{Energy and System Size Dependence of Charged Hadron Transverse Momentum 
Spectra from Cu+Cu and Au+Au Collisions at $\sqrt{s_{_{\it NN}}} =$~62.4 and 200 GeV} 

\authors{ 
{Edward Wenger$^1$ for the PHOBOS Collaboration %
\index{Wenger, E. (PHOBOS)} 
}\\[2.812mm]
{\normalsize
\hspace*{-8pt}$^1$ Massachusetts Institute of Technology,\\ 
Cambridge, MA 02139-4307, USA\\[0.2ex] 
}}
 
\abstract{The PHOBOS collaboration has measured transverse momentum 
distributions of charged hadrons produced in Cu+Cu collisions at 
$\sqrt{s_{_{\it NN}}} =$~200 and 62.4~GeV.  
The nuclear modification factor $R_{\it AA}^{Npart}$ is calculated relative to p+p
data at both collision energies as a function of collision centrality.  
For the same number of participating nucleons, $R_{\it AA}^{Npart}$ is essentially
the same in both systems over the full range of $p_{T}$ that is measured.  
In addition, we observe that within experimental uncertainties, the ratio of 
200 GeV to 62.4 GeV Cu+Cu yields has only a moderate centrality dependence and 
is consistent with the value previously measured in Au+Au collisions for 
a broad range of $p_{T}$. }

 
\makeindex
\begin{document}
 
\maketitle

\section{Introduction}\label{intro}
The yield of charged hadrons produced in Cu+Cu collisions at 
energies of $\sqrt{s_{_{\it NN}}} =$~200 and 62.4~GeV has been measured
with the PHOBOS detector at the Relativistic Heavy Ion Collider (RHIC) at 
Brookhaven National Laboratory. The data are presented as a function of 
transverse momentum ($p_T$) and collision centrality.  The goal of these
measurements is to study the modification of particle production compared 
to nucleon-nucleon collisions at the same energy.

This measurement was motivated by the results from Au+Au collisions for 
$\sqrt{s_{_{\it NN}}} =$~62.4 and 200~GeV.  Hadron production at these energies 
was found to be strongly suppressed relative to expectations based on 
an independent superposition of  nucleon-nucleon collisions at $p_T$ of 
2--10 GeV/c \cite{phenix_quench,phenix_highpt_npart,star_highpt_npart,phobos_highpt_npart}. 
Such a suppression had been predicted to occur as a consequence of the
energy loss of high-$p_T$ partons in the dense medium formed in Au+Au
collisions \cite{jet_quench_theory}.  This hypothesis is consistent with
the observed absence of this effect in deuteron--gold collisions at the
same collision energy \cite{phenix_dAu,star_dAu,brahms_dAu,phobos_dAu}.        
The results presented here for Cu+Cu collisions at $\sqrt{s_{_{\it NN}}}
=$~200 and 62.4~GeV bridge the gap between the Au+Au and d+Au systems, 
allowing a unique examination of the dependence of high $p_T$ suppression 
on system size.  

Furthermore, it has been shown for Au+Au collisions that, within
experimental uncertainty, a surprisingly clean factorization of the energy 
and centrality dependence of charged hadron yields exists at all measured
values of $p_{T}$ \cite{phobos_highpt_62}.  The Cu+Cu system allows us to 
explore whether this factorization persists down to a system of 20 participant 
nucleons or whether there is a critical size where this factorization begins to
break down.  
 
\section{Technical details}\label{techno}  
The data were collected using the PHOBOS two-arm magnetic spectrometer 
\cite{phobos_nim}.
The primary event trigger used the time difference between signals in 
two sets of 10 \v Cerenkov counters, located at $4.4< | \eta | <4.9$, to select
collisions along the beam-axis that were close to the nominal vertex position.
For the analysis presented here, events were divided into centrality classes
based on the total energy deposited in the octagon silicon detector, covering 
pseudo-rapidities $ |\eta |< 3.0$.   
A full detector simulation using HIJING events \cite{phobos_cent_200,hijing} 
was used to estimate $\langle N_{part} \rangle$ for each centrality class, 
and the corresponding $\langle N_{coll} \rangle$ values were obtained from 
a Monte Carlo Glauber calculation \cite{phobos_cent_200}. 

\section{Results}\label{result}
\begin{figure}[htb]
\centerline{
  \mbox{\includegraphics[width=11cm]{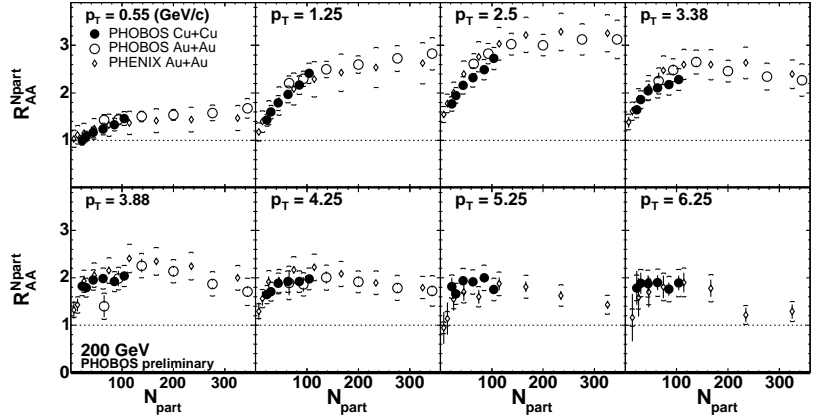}}}
\vspace{-0.35cm}
\caption{ \label{RAA200vsNpart}
  Nuclear modification factor, $R_{\it AA}^{Npart}$ versus $N_{part}$ in bins of 
  $p_{T}$ at $\sqrt{s_{_{\it NN}}} =$ 200~GeV for Cu+Cu (filled symbols) and 
  Au+Au (open symbols).  Systematics are shown with brackets (90\% C.L.).  
  Data from the PHENIX collaboration are also included (open diamonds) where 
  they have been published \cite{phenix_highpt_npart}.}
\end{figure}

\begin{figure}[htb]
\centerline{
  \mbox{\includegraphics[width=11cm]{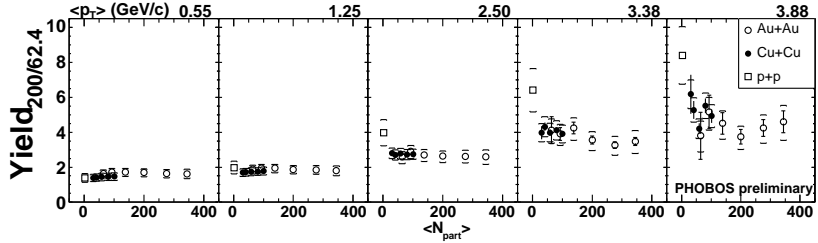}}}
\vspace{-0.25cm}
\caption{ \label{200over62}
  Ratio of yields at $\sqrt{s_{_{\it NN}}} =$ 200~GeV to 62.4 GeV for Cu+Cu 
  (filled circles) and Au+Au (open circles).  The ratio of yields in elementary 
  nucleon-nucleon collisions is also shown (open squares) \cite{pp_reference}.}
\end{figure}

It has been previously noted that the observed strong centrality dependence of
$R_{\it AA}$ at $\sqrt{s_{_{\it NN}}} =$ 200~GeV corresponds to a relatively 
small change in the yield per participating nucleon \cite{phobos_highpt_npart}
and that over the same centrality range, the total yield of charged particles
per participating nucleon is constant within experimental uncertainties
\cite{universality}.  These observations lead us to define $R_{\it AA}^{Npart}$ 
in analogy to $R_{\it AA}$, where now we scale the reference spectrum by 
$N_{part}/2$ rather than $N_{coll}$.
\begin{equation}
R_{\it AA} = \frac{\sigma_{pp}^{inel}}{\langle N_{\it coll} \rangle} 
              \frac{d^2 N_{\it AA}/dp_T d\eta} {d^2 \sigma_{pp}/dp_T d\eta},
\hspace{1.0cm}
R_{\it AA}^{Npart} = \frac{\sigma_{pp}^{inel}}{\langle N_{\it part} \rangle/2} 
              \frac{d^2 N_{\it AA}/dp_T d\eta} {d^2 \sigma_{pp}/dp_T d\eta}.
\end{equation}
In Fig.~\ref{RAA200vsNpart} we explore the centrality dependence of 
$R_{\it AA}^{Npart}$ at $\sqrt{s_{_{\it NN}}} =$ 200~GeV in detail.  Over 
the range of $p_{T}$ that we measure,
the bulk particle production seems to depend only on the size of the system, 
that is the Cu+Cu and Au+Au spectra look identical for the same number of 
participating nucleons.  
This observation appears to hold at $\sqrt{s_{_{\it NN}}} =$ 62.4~GeV as well, 
as is shown implicitly in Fig.~\ref{200over62} where the ratios of spectra at 
two energies again coincide for the same system-size.

In Fig.~\ref{200over62} the ratio of yields in 200 GeV to 62.4 GeV is plotted 
versus centrality for a range of transverse momenta.  Within our experimental
uncertainties, it is shown that the ratio of yields in Cu+Cu has only a 
moderate centrality dependence and is consistent with the Au+Au value for all
measured $p_{T}$.

\vspace{-0.5cm}

\section{Conclusions}\label{concl}
Particle production at $p_{T} > 1$~GeV/c in heavy-ion collisions is expected
to be influenced by the interplay of many effects.  This includes 
$p_{T}$-broadening due to initial and final state multiple scattering (the 
`Cronin effect'), the medium-induced energy loss of fast partons, as well as 
the effects of collective transverse velocity fields and of parton recombination
\cite{theory_review}.  Considering the significantly different geometries of
Au+Au and Cu+Cu collisions with the same number of participant nucleons, it 
is not obvious, {\it a priori}, that these effects should conspire to give the 
same spectra in both systems over the measured $p_{T}$ range.    

Moreover, the results presented here demonstrate, within experimental 
uncertainty, a surprisingly clean factorization of the energy and centrality
dependence of charged hadron yields that persists for very small systems at all
measured values of $p_{T}$.  
This factorization  of energy and centrality is also a characteristic feature of
the total and differential particle yields \cite{phobos_lim_frag,universality} 
and of multi-particle correlation measurements such as Bose-Einsten 
correlations \cite{hbt}.  These observations provide further evidence of global
constraints on particle production and appear to be a key feature of heavy-ion 
collisions that remains to be understood.

\section*{Acknowledgments}
We acknowledge the generous support of the Collider-Accelerator Department.
This work was partially supported by U.S. DOE grants 
DE-AC02-98CH10886,
DE-FG02-93ER40802, 
DE-FC02-94ER40818,  
DE-FG02-94ER40865, 
DE-FG02-99ER41099, and
W-31-109-ENG-38, by U.S. 
NSF grants 9603486, 
0072204,            
and 0245011,        
by Polish KBN grant 1-P03B-062-27(2004-2007),
by NSC of Taiwan Contract NSC 89-2112-M-008-024, and
by Hungarian OTKA grant (F 049823).

\vfill\eject
\end{document}